\documentclass[preprint,12pt]{elsarticle}

\journal{Astroparticle Physics}

\begin{document}

\begin{frontmatter}

\title{Nuclear enhancement factor in calculation of Galactic diffuse
gamma-rays: A new estimate with DPMJET-3}

\author{Masaki Mori*%
\footnote[1]{*Present address: Department of Physics, College of Science and Engineering, Ritsumeikan University, Noji Higashi 1-1-1, Kusatsu, Shiga 525-8577, Japan}}

\ead{morim@icrr.u-tokyo.ac.jp, morim@fc.ritsumei.ac.jp}

\address{Institute for Cosmic Ray Research, University of Tokyo, \\
5-1-5 Kashiwanoha, Kashiwa, Chiba 277-8582, Japan}

\begin{abstract}
A new calculation of nuclear enhancement factor, used in
estimation of Galactic diffuse gamma-ray flux from proton-proton
interaction in order to take account of heavy nuclei included
in cosmic-rays and interstellar matter, 
is presented by use of a Monte Carlo simulator,
DPMJET-3. A new value of 1.8--2.0 in the energy range
of 6--1000 GeV/nucleon, slightly increasing with
kinetic energy of projectile cosmic rays, is about 20\% 
larger than previous estimates.
\end{abstract}


\end{frontmatter}

\section{Introduction}

Majority of cosmic gamma-rays observed by EGRET onboard the
Compton gamma-ray observatory concentrate around the
Galactic plane, and they are known as Galactic 
diffuse gamma-rays \cite{Fic97}. 
Gamma-ray point sources near the Galactic plane
are observed above this diffuse `background' gamma-rays,
and the diffuse emission model based on high-energy
particle interaction with interstellar matter/field
is extremely important for the analysis of gamma-ray
emitting objects in the Universe.

The dominant component of Galactic diffuse gamma-rays 
above 100 MeV is well explained by decay gamma-rays from 
neutral pions generated in nuclear interaction
of high-energy cosmic-rays and interstellar matter
\cite{Ber93,Hun97}.
In calculation of this process,
most authors have computed gamma-ray flux generated
by neutral pion decay which is produced in
high-energy proton interaction on proton target
(see Kamae et al.~\cite{Kam06} for the most recent
calculation).
Actually, there are heavy nuclei both in cosmic-rays
and target matter. This fact is taken into account
by introducing a {\it nuclear enhancement factor},
 $\epsilon_{\rm M}$,
to be multiplied to the result of gamma-ray yield
assuming cosmic-ray protons on proton target only.
Previous estimates are as follows:
$\sim 2.0$ by Stecker \cite{Ste70},
 1.5 by Cavallo and Gould \cite{Cav71}, 
$1.6\pm0.1$ by Stephens and Badhwar \cite{Ste81}, 
1.45 by Dermer \cite{Der86a}, 
$1.52+\max(1,(T_p/{\rm 100\,GeV})^{0.12})$ by Mori \cite{Mor97},
where $T_p$ stands for kinetic energy of cosmic-ray proton.
Mori \cite{Mor97} used the low-energy value of 1.52
given by Gaisser and Shaefer \cite{Gai92} derived
for calculation of cosmic antiproton flux without a factor
related only to propagation, and took the spectral difference
of heavier nuclei into account.

Later, Simon, Molner and Roesler \cite{Sim98} calculated
interstellar secondary antiproton flux with a Monte Carlo model,
DTUNUC, to study contributions to the antiproton flux
from $p$-nucleus, nucleus-$p$ and nucleus-nucleus collisions.
They gave the energy dependence on the nuclear enhancement factor
(which they call `nuclear scaling factor' in their paper)
which differs significantly in low-energy region ($T_{\bar{p}}<10$\,GeV)
from a constant value as given by Gaisser and Shaefer \cite{Gai92}.
The situation is similar to the case of Galactic diffuse gamma-rays,
and we need detailed calculation of nuclear effect
taking account of energy dependence for the new era of gamma-ray
astronomy with a launch of {\it Fermi} (formerly called {\it GLAST}) 
gamma-ray space telescope in June 2008 \cite{Mic07}.

Here we use a Monte Carlo model, DPMJET-3 \cite{Roe01},
 which is an updated
version of DTUNUC, to calculate a nuclear enhancement factor
for the Galactic diffuse gamma-ray flux.%
\footnote{Recently, Huang et al.\  calculated gamma-ray yield from 
(p, $\alpha$)+(interstellar matter) interactions using the same 
Monte Carlo model \cite{Hua07}, but their results are given as final 
spectra and the nuclear effect on the gamma-ray production was 
not explicitly discussed in their paper.}

\section{Calculation}

We calculated gamma-ray yield from 
$p$-nucleus, nucleus-$p$ and nucleus-nucleus collisions
using DPMJET-3.
It is a Monte Carlo model capable of 
simulating hadron-hadron, hadron-nucleus, nucleus-nucleus, 
from a few GeV up to the highest cosmic-ray energies,
and has been successfully applied to the description 
of hadron production in high-energy collisions \cite{Roe01}.
 To check our calculation with this model,
``multiplication factor'', $m_{ip}$ and $m_{i\alpha}$
for proton and Helium target, respectively,
given in Gaisser and Shaefer \cite{Gai92}, was recalculated
for $T=10$~GeV where $T$ is a kinetic energy per nucleon.
This is a relative yield of gamma-rays
from nucleus-$p$ and nucleus-Helium collisions 
compared with that from $p$-$p$ collisions,
and is defined as $m_{ip} \equiv (\sigma_{ip}/\sigma_{pp})
(\langle n_{\gamma,ip}\rangle/\langle n_{\gamma,pp}\rangle)$
and similarly for $m_{i\alpha}$, where $\sigma_{ip}$ ($\sigma_{pp}$)
and $\langle n_{\gamma,ip}\rangle$ ($\langle n_{\gamma,pp}\rangle$) are 
a production cross section and an average number
of gamma-rays produced  in nucleus-$p$ ($p$-$p$) collision, respectively.
The results are given in Table \ref{tab:mf}.
For each entry, 20,000 collisions were generated.
Reasonable agreement is seen, and in fact, using the flux values
at 10 GeV/nucleon, nuclear enhancement factor can be calculated as
\begin{equation}
\epsilon_{\rm M}=1+\sum_i m_{ip}{\phi_i(T) \over \phi_p(T)}
+\sum_i m_{i\alpha}{\phi_i(T) \over \phi_p(T)}
\times {r \over 1-r} = 1.52
\label{eq:b}
\end{equation}
and is close to 1.52, the value given by
Gaisser and Schaefer \cite{Gai92}, 
assuming the interstellar matter abundance
and cosmic-ray composition given by ref.\cite{Gai92}
(with the fraction of Helium
in interstellar matter $r=0.07$).
Thus the low-energy value of the nuclear enhancement factor
changes little by use of modern nuclear interaction model.

\begin{table}
\begin{center}
\caption{Multiplication factor at $T$=10~GeV/nucleon. G\&S is quoted from 
ref.\protect\cite{Gai92}.}
\label{tab:mf}
\begin{tabular}{l|cc|cc}\hline
Nuclei         &\multicolumn{2}{|c|}{$m_{ip}$} 
&\multicolumn{2}{|c}{$m_{i\alpha}$} \\
               & DPMJET-3 & G\&S & DPMJET-3 & G\&S \\ \hline 
H ($A$=1)      & (1) & (1)       & 3.81 & 3.57 \\
He ($A$=4)     & 3.68 & 3.57     & 14.2 & 12.6 \\
CNO ($A$=14)   & 11.7 & 11.4 (N) & 42.5 & 40 (N) \\
Mg-Si ($A$=25) & 20.3 & 20 (Al)  & 73.2 & 71 (Al) \\
Fe ($A$=56)    & 38.8 & 40       & 142 & 135 \\ \hline 
\end{tabular}
\end{center}
\end{table}

\begin{figure}
\begin{center}
  \includegraphics[height=90mm]{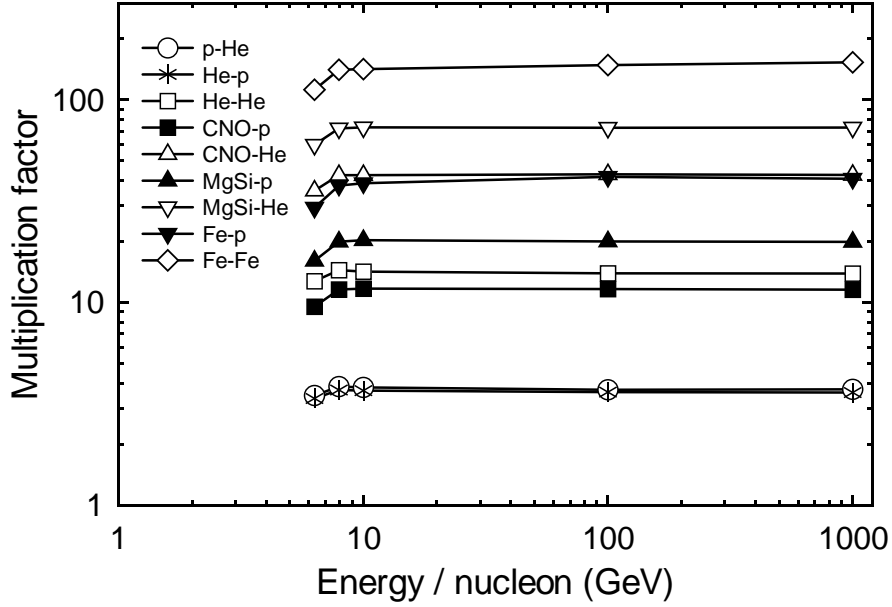}
  \caption{Energy dependence of multplification factor for 
  various combination of projectile-target nuclei.}
  \label{fig:mf-edep}
\end{center}
\end{figure}

Figure \ref{fig:mf-edep} shows the energy dependence of
multiplication factor.
Due to limitations of the DPMJET-3 model, calculations
below about 5 GeV/nucleon produce errors and we stop
calculations there.
Table \ref{tab:mf-h} shows multiplication factors for heavy 
target nuclei calculated using DPMJET-3. 
Note the approximate symmetry between 
projectile and target nuclei (ex. $p$-Fe vs. Fe-$p$).

\begin{table}
\begin{center}
\caption{Multiplication factor at $T$=10~GeV/nucleon for heavy nuclei target.}
\label{tab:mf-h}
\begin{tabular}{l|ccc}\hline
Projectile     &\multicolumn{3}{|c}{Target} \\ 
nuclei         & CNO   & MgSi  & Fe    \\ \hline 
H ($A$=1)      & 11.6  & 20.1  & 38.9  \\
He ($A$=4)     & 42.3  & 73.4  & 143.1 \\
CNO ($A$=14)   & 120.6 & 204.2 & 386.3 \\
Mg-Si ($A$=25) & 204.2 & 343.1 & 628.8 \\
Fe ($A$=56)    & 384.4 & 634.0 & 1067 \\ \hline 
\end{tabular}
\end{center}
\end{table}

Next, we explore other elements of nuclear enhancement
factor: cosmic-ray spectra and composition of interstellar matter.

The primary cosmic-ray spectra, $\phi_i(T)$, 
are summarized in empirical formulas by Honda, 
Kajita, Kasahara and Midorikawa (2004) \cite{Hon04}:
\begin{equation}
 \phi_i(T)= K(T+b \exp(-c\sqrt{T}))^{-\alpha}
 \,\,\,[{\rm m}^{-2}{\rm s}^{-1}{\rm sr}^{-1}]
\label{eq:a}
\end{equation}
with parameters
$\alpha$, $K$, $b$ and $c$ tabulated in Table \ref{tab:para},
where cosmic-ray nuclei are divided into five groups.
If we use these spectra, the nuclear enhancement factor
is calculated to be 1.66 at 10 GeV/nucleon, about 10\% larger than
that by Ref.\cite{Gai92}.

\begin{table}
\begin{center}
\caption{Primary cosmic-ray spectral parameters \cite{Hon04}.}
\label{tab:para}
\begin{tabular}{l|cccc}\hline
Component      & $\alpha$ & $K$ & $b$ & $c$ \\ \hline
H ($A$=1)      & $2.74\pm0.01$ & $14900\pm600$ & 2.15 & 0.21 \\
He ($A$=4)     & $2.64\pm0.01$ & $600\pm30$   & 1.25 & 0.14 \\
CNO ($A$=14)   & $2.60\pm0.07$ & $33.2\pm5$  & 0.97 & 0.01 \\
Mg-Si ($A$=25) & $2.79\pm0.08$ & $34.2\pm6$  & 2.14 & 0.01 \\
Fe ($A$=56)    & $2.68\pm0.01$ & $4.45\pm0.50$  & 3.07 & 0.41 \\ \hline 
\end{tabular}
\end{center}
\end{table}

As for the composition of interstellar medium, 
Meyer summarizes elemental abundances in
local Galactic interstellar media in detail \cite{Mey85}.
In his Table 2 the He/H ratio is 0.096,%
\footnote{This ratio is also consistent with the average
of abundance in interstellar gas in various objects
tabulated in Table 21.9
of Ref.\cite{Cox00}, $0.096\pm0.07$ (omitting 30 Dor and Crab
which shows extra values).} which is
significantly larger than $0.07/0.93=0.075$ assumed
in Gaisser and Shaefer \cite{Gai92}.
This fact leads to a further larger value of $\epsilon_{\rm M}$.
Numerically, if we use 
relative abundance of H$:$He$:$CNO$:$NeMgSiS$:$Fe$=1:0.096:
1.38\times10^{-3}:2.11\times10^{-4}:3.25\times10^{-5}$
following the compilation by Meyer \cite{Mey85}.
Then we obtain the nuclear enhance factor at 10~GeV/nucleon:
\begin{eqnarray}
\epsilon_{\rm M}&=& 1+\epsilon({\rm H})+\epsilon({\rm He})
+\epsilon({\rm CNO})+\epsilon({\rm NeMgSiS})+\epsilon({\rm Fe}) =1.84
\label{eq:c}
\end{eqnarray}
where $\epsilon(i)$ is the contribution from interstellar
nuclei $i$ bombarded by cosmic-rays with the composition
given by the flux expressed by Eq.(\ref{eq:a}), except $p$-$p$ interaction,
calculated in the same manner as in Eq.(\ref{eq:b}).
This value is about 20\% higher than previous calculations.
As discussed above, most of the difference resides in abundance 
of Helium in cosmic-ray flux and interstellar medium.
Table \ref{tab:nef-comp} shows the contribution of each component
to the nuclear enhancement factor.

\begin{figure}
\begin{center}
  \includegraphics[height=90mm]{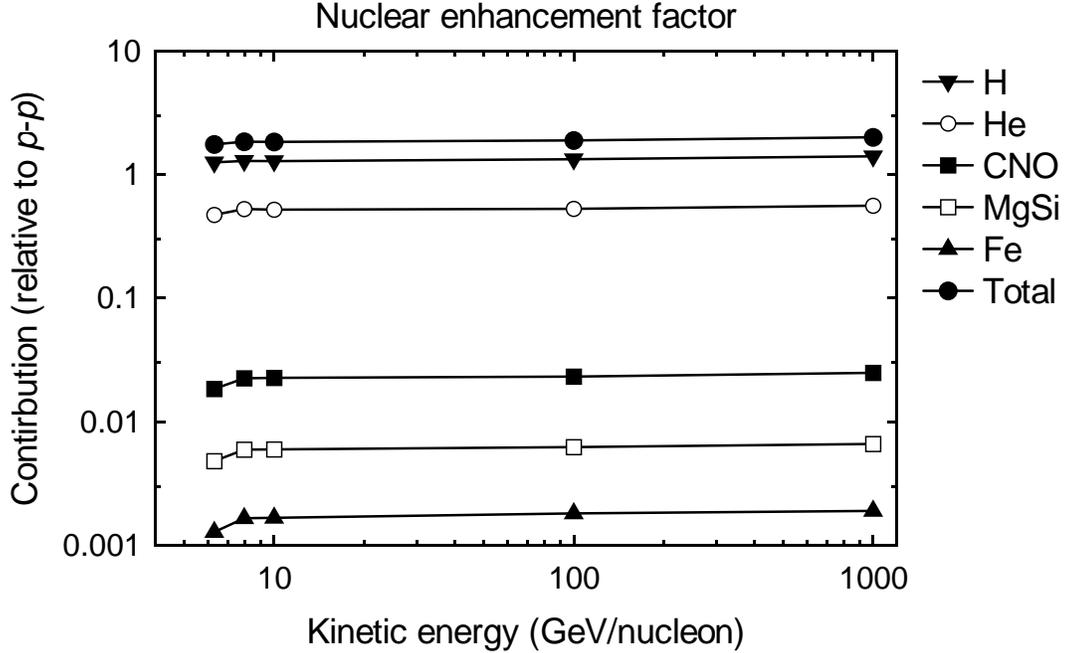}
  \caption{Energy dependence of nuclear enhancement factor contributions 
  from each ISM component.}
  \label{fig:nef}
\end{center}
\end{figure}

\begin{table}
\begin{center}
\caption{Nuclear enhancement factor decomposed to each component
at 10 GeV/nucleon.}
\label{tab:nef-comp}
\begin{tabular}{l|ccccc|c}\hline
  & \multicolumn{5}{|c|}{Target}&Sum \\
  & H & He & CNO & NeMgSiS & Fe  \\ \hline
Projectile & & & & & \\ 
\hspace{1em}H   & 1     & 0.405 & 0.0177 & 0.0047 & 0.0006 & 1.428 \\
\hspace{1em}He  & 0.203 & 0.083 & 0.0036 & 0.0035 & 0.0004 & 0.293 \\
\hspace{1em}CNO & 0.038 & 0.015 & 0.0006 & 0.0018 & 0.0002 & 0.055 \\
\hspace{1em}MgSi& 0.033 & 0.013 & 0.0005 & 0.0026 & 0.0003 & 0.049 \\
\hspace{1em}Fe  & 0.014 & 0.006 & 0.0002 & 0.0021 & 0.0002 & 0.022 \\ \hline 
Sum             & 1.288 & 0.520 & 0.023 & 0.0147 & 0.0017 & 1.845 \\ \hline 
\end{tabular}
\end{center}
\end{table}

Fig.~\ref{fig:nef} shows the energy dependence of the nuclear
enhancement factor including contributions from each ISM
abundance. The total factor is also tabulated in Table~\ref{tab:nef}.
Although it is nearly constant above about 10 GeV/nucleon,
it decrease gradually toward lower energies.
However, the calculation is limited above above 5 GeV/nucleon,
again due to the limitation of DPMJET-3.

\begin{table}
\begin{center}
\caption{Energy dependence of nuclear enhancement factor.}
\label{tab:nef}
\begin{tabular}{l|ccccc}\hline
Kinetic energy (GeV/nucleon) & 6.31 & 7.94 & 10.0 & 100 & 1000  \\ \hline
Nuclear enhancement factor   & 1.75 & 1.84 & 1.84 & 1.89 & 2.00 \\ \hline
\end{tabular}
\end{center}
\end{table}

\section{Conclusion}

Nuclear enhancement factor used in
models of the Galactic diffuse gamma-rays from cosmic-ray and 
interstellar matter
interaction has been calculated by use of a Monte Carlo simulator,
DPMJET-3. A new value of 1.8--2.0, slightly increasing with
kinetic energy of projectile cosmic rays in the range 6--1000~GeV/nucleon, 
is about 20\% larger than previous estimates.
This factor could have a larger energy dependence below
5 GeV/nucleon where DPMJET-3 is not applicable
(see Simon et al.\ \cite{Sim98} for the case of antiproton production),
but we do not have a proper tool to treat that energy region.

\medskip
\noindent{\bf Acknowledement}

The author wish to thank Dr.~S.~Roesler for his kind approval of use
and supply of the DPMJET-3 code.

\end{document}